\documentclass[12pt]{article}
\usepackage{latexsym}
\usepackage{graphicx}

\thicklines                         
\long \def \blockcomment #1\endcomment{}

\begin{document}           
\baselineskip=0.33333in

\vglue 0.5in
\begin{center}{\bf Physical Consequences of  \\
Mathematical Principles}
\end{center}
\begin{center}E. Comay$^*$
\end{center}

\begin{center}
Charactell Ltd. \\
PO Box 39019 \\
Tel-Aviv, 61390, Israel
\end{center}
\vglue 0.5in
\noindent
PACS No: 03.30.+p, 02.30.Xx, 11.30.Cp, 03.50.De
\vglue 0.2in
\noindent
Abstract:

Physical consequences are derived from the following mathematical
structures: the variational principle,
Wigner's classifications of the irreducible representations of the Poincare
group and the duality invariance of the homogeneous Maxwell equations.
The analysis is carried out within the validity domain of special
relativity. Hierarchical relations between physical theories are used.
Some new results are pointed out together with their comparison
with experimental data. It is also predicted that a genuine
Higgs particle will not be detected.
\newpage
\noindent
{\bf 1. Introduction}
\vglue 0.33333in

Physics aims to describe processes which are observed in
the real world. For this purpose,
mathematical formulations of physical theories are constructed.
Mathematical elements of a physical theory can be divided into three sets:
elements that play a relative fundamental role and are regarded as
cornerstones of the theory's structure,
elements used as a derivation tool and final formulas that describe
the behavior of a given system. This kind of classification is used
here for the convenience of the presentation. In particular, what is
regarded here as a fundamental element may, in principle, be derived
from more profound mathematical elements.

This work regards the following mathematical structures as
cornerstones of the discussion. The variational principle and its
relevant Lagrangian density; Wigner's analysis of the
irreducible representations
of the Poincare group; the duality invariance of the homogeneous
Maxwell equations. Some well known results of these elements are
pointed out alongside others that are not very well known.
Boldface numbers
are used for marking the latter kind of results. It is shown that
some of these results fit experimental data whereas others are used
as a prediction of yet unknown experimental data.

The discussion is carried out within a framework that is based
on the following theoretical elements. First, Special Relativity
is regarded as a covering theory and all expressions must be
consistent with relativistic covariance. The De Broglie relation
between the particle's wave properties and its energy-momentum
is used. Another issue is related
to the hierarchical relations between physical theories.
(A good discussion of this issue can be found in [1], pp. 1-6.)
The following lines explain this issue in brief.

Every physical theory applies to a limited set of processes.
For example, let us take the problem of moving bodies.
It is well known that physical theories yield very good
predictions for the motion of planets around the sun. On the other hand,
nobody expects that a physical theory be able to predict the specific
motion of an eagle flying in the sky. This simple example proves that
the validity of a physical theory should be evaluated only with respect
to a limited set of experiments. The set of experiments which can
be explained by a physical theory is called its domain of validity.
The relations between domains of validity define
hierarchical relations between the corresponding theories.
For example, given theories $A,\;B$ and $A$'s domain of validity
is a subset of $B$'s domain of validity then $B$'s rank is higher
than that of $A$.

An examination of Newtonian mechanics and relativistic mechanics
illustrates the notion of hierarchical relations between theories.
Newtonian mechanics is good for
low velocity experiments (because its predictions are consistent with the
error range of measurements). On the other hand, relativistic mechanics
is good even for velocities that approach the speed of light.
Two conclusions can be derived from these properties of the theories:
First, relativistic mechanics has
a more profound basis because it is valid for all
experiments where Newtonian mechanics holds {\em and}
for many other experiments where Newtonian mechanics
fails. Another aspect of the relations between Newtonian mechanics
and relativistic mechanics is that Newtonian mechanics imposes
constraints on the form of the low velocity limit of relativistic mechanics.
Indeed, the low velocity limit of relativistic mechanics is
(and must be) consistent
with Newtonian formulas. Below, this kind of constraint is called
{\em constraint imposed by a lower rank theory}. Some of the theoretical
derivations included below rely on this principle.

The Lorentz metric used is diagonal and its entries are (1,-1,-1,-1).
Greek indices run from 0 to 3.
Expressions are written in units where $\hbar = c = 1$. In this
system of units there is just one dimension. Here it is taken to be
that of length. Therefore, the dimension of a physical quantity is a
power of length and is denoted by $[L^n]$. In particular, energy and
momentum take the dimension $[L^{-1}]$.
The symbol $Q_{,\mu}$
denotes the partial derivative of the quantity $Q$ with respect to
$x^\mu $. An upper dot denotes a differentiation with respect to time.

The second section discusses quantum mechanical consequences of the
variational principle. The Dirac equation is examined in the third section.
The fourth section shows inconsistencies of
the Klein-Gordon (KG) and the Higgs equations.
The fifth section examines results obtained from Wigner's classification
of the irreducible representations of the Poincare group. Consequences of
the duality invariance of the homogeneous Maxwell equations
together a regular charge-monopole theory
are discussed in the sixth section.
The seventh section contains concluding remarks.

\vglue 0.66666in
\noindent
{\bf 2. The Variational Principle}
\vglue 0.33333in

This section is dedicated to the form of a quantum theory of a {\em massive
particle}. Let us examine the pattern obtained in a two slit interference
experiment. Here one finds bright and dark strips. A completely dark
interference
point indicates that a full anti-phase destruction takes place there.
Obviously, this property should be obtained in every Lorentz frame of
reference. It follows that the phase must depend on a Lorentz scalar.

The quantity which is suitable for this purpose is the action of the
system. Thus, let us examine a Lagrangian density of the system and
its action
\begin{equation}
S = \int {\mathcal L}(\psi ,\psi _{,\mu}) d^4x^\mu .
\label{eq:ACTION}
\end{equation}
Now, if the Lagrangian density is a Lorentz scalar then also the action
is a Lorentz scalar. Therefore, it is concluded that

\begin{itemize}
\item[{1.}] A relativistically consistent quantum theory may be
derived from a Lagrangian density which is a Lorentz scalar.
\end{itemize}

Another issue is related to the dimension of the quantities. The phase
is an argument of an exponent. Therefore, it must be dimensionless. Thus,
in the system of units used here the action is dimensionless and
satisfies this requirement. It follows that
\begin{itemize}
\item[{\bf {2.}}] An acceptable Lagrangian density must have the
dimension $[L^{-4}]$.
\end{itemize}
{\em This conclusion means that the wave function $\psi $ acquires a
well defined dimension.}

(Remark. The foregoing arguments indicate that if one wishes to take
an alternative way for constructing a
relativistically self-consistent quantum theory,
then one must find another physically meaningful quantity that is
a dimensionless Lorentz scalar and is
suitable for taking the role of the particle's phase. Apparently, such a
quantity does not exist. If this claim is correct then the variational
principle is also a necessary condition for constructing a self-consistent
relativistic quantum theory.)

Another point is related to the independent variables $x^\mu $
of the wave function
\begin{equation}
\psi(x^\mu )
\label{eq:PSI}
\end{equation}
which is a {\em single set} of four space-time coordinates. Therefore
$(\!\!~\ref{eq:PSI})$ cannot describe a composite particle, because
such a particle requires, besides a description of
the space-time location of its
center of energy, additional coordinates for describing its internal
structure. Therefore,
\begin{itemize}
\item[{\bf {3.}}] The wave function $\psi(x^\mu )$ describes an
elementary structureless pointlike particle.
\end{itemize}
This result is consistent with the nature of an elementary classical
particle (see [2], pp. 46,47). Below it is applied as a useful criterion
for evaluating experimental data.

The Lagrangian density is used here as the cornerstone of the theory.
Hence, the particle's equations of motion are the corresponding
Euler-Lagrange equations (see [3], p. 14; [4], p. 16)
\begin{equation}
\frac {\partial }{\partial x^\mu} \frac {\partial {\mathcal L}}
{\partial \frac {\partial \psi}{\partial x^\mu }} -
\frac {\partial {\mathcal L}}{\partial \psi} = 0.
\label{eq:ELEQ}
\end{equation}
On this basis it is concluded that
\begin{itemize}
\item[{{4.}}] The particle's equations of motion are the Euler-Lagrange
equations derived from the Lagrangian density.
\end{itemize}
Obviously, different kinds of Lagrangian density yield different
equations of motion. This point is discussed later.

Another issue is the consistency of a quantum theory of a massive
particle with the classical theory, where the latter
provides an example of
constraints imposed by a lower rank theory. The classical limit
of quantum mechanics
is discussed in the literature (see [5], pp.19-21 and elsewhere;
[6], pp. 25-27, 137-138).

In order to do that,
the quantum theory should provide expressions for the energy and the
momentum of the particle. As a matter of fact, having an appropriate
expression for the energy at the system's rest frame is enough. Indeed,
a Lorentz boost guarantees that the theory provides appropriate
expressions for the energy and momentum in any reference frame.
Therefore, the following lines examine
the construction of an expression for the energy of a massive quantum
mechanical particle {\em in its rest frame}.
For this end, let us take the Lagrangian density and
construct the following second rank tensor
(see [4], p. 19)
\begin{equation}
{\mathcal T}_{\mu \nu} =
\frac {\partial {\mathcal L}}
{\partial \frac {\partial \psi}{\partial x_\mu }}
\frac {\partial \psi}{\partial x^\nu }  -
{\mathcal L}g_{\mu \nu }.
\label{eq:EMTENSOR}
\end{equation}

Now, density is a 0-component of a 4-vector and the same is true for
energy. Hence, energy density is a (0,0) component of a second rank
tensor. Moreover, like the dimension of the Lagrangian density,
the dimension of ${\mathcal T}_{\mu \nu}$ of
$(\!\!~\ref{eq:EMTENSOR})$ is $[L^{-4}]$. This is also the dimension
of energy density. Now, in quantum mechanics, the Hamiltonian is
regarded as the energy operator.
Thus, the entry ${\mathcal T}_{00}$ of
$(\!\!~\ref{eq:EMTENSOR})$ is regarded as an expression for
the Hamiltonian density
\begin{equation}
{\mathcal H} = \dot {\psi} \frac {\partial {\mathcal L}}
{\partial \dot {\psi}} - {\mathcal L}.
\label{eq:HDENSITY}
\end{equation}

It is explained below why an expression for density is required.
Here, density properties
can be readily taken from electrodynamics (see [2], pp. 73-75).
Density must have the dimension $[L^{-3}]$ and be a 0-component of
a 4-vector satisfying the continuity equation
\begin{equation}
j^\mu _{,\mu } = 0.
\label{eq:CONTINUITY}
\end{equation}

At this point, one may take either of the following alternatives:

\begin{itemize}
\item[{A.}]
Use the Hamiltonian density ${\mathcal H}$ together
with the density expression and extract the Hamiltonian
differential operator $H$, operating on $\psi $. The energy is an
eigenvalue of this operator:
\begin{equation}
H\psi = E\psi,
\label{eq:HPSIE}
\end{equation}
Now the De Broglie relation
\begin{equation}
i \frac {\partial \psi}{\partial t} = E\psi,
\label{eq:EPSI}
\end{equation}
yields the differential equation
\begin{equation}
i \frac {\partial \psi}{\partial t} = H\psi.
\label{eq:HPSI}
\end{equation}
At this point one can construct a Hilbert space that includes all
eigenfunctions of the Hamiltonian $H$.

\item[{B.}] Use the expression for density as
an inner product for $\psi $ and construct an orthonormal basis for
the corresponding Hilbert space. Next construct the Hamiltonian
matrix. For the $i,j$ functions of the Hilbert space
basis, the Hamiltonian matrix
element is
\begin{equation}
H_{ij} = \int {\mathcal H} (\psi _i, \psi _{i,\mu },\psi _j ,\psi _{j,\nu })
d^3x
\label{eq:HIJ}
\end{equation}
At this point, the Hamiltonian matrix is diagonalized and
energy eigenfunctions and eigenvalues are obtained.

\end{itemize}
Obviously, the mathematical structures of A and B are relevant to the
same data. Therefore, both methods construct one and the same Hilbert
space.

Equation $(\!\!~\ref{eq:HPSI})$ makes the following problem.
As stated above, the Euler-Lagrange equation $(\!\!~\ref{eq:ELEQ})$
is the system's equation of motion. On the other hand,
$(\!\!~\ref{eq:HPSI})$ is {\em another} differential equation. Hence,
the following requirement should be satisfied.
\begin{itemize}
\item[{\bf {5.}}] Requirement 1:
The first order differential equation $(\!\!~\ref{eq:HPSI})$
should be consistent with the Euler-Lagrange equation of the theory
$(\!\!~\ref{eq:ELEQ})$.
\end{itemize}

The next two sections are devoted to two specific kinds of
Lagrangian density of massive particles.

\vglue 0.66666in
\noindent
{\bf 3. The Dirac Field}
\vglue 0.33333in

It is shown here that the Dirac field satisfies the
requirements derived above and that experimental data support the theory.
The formulas are written in the standard notation [3,7].

The Dirac Lagrangian density is
\begin{equation}
{\mathcal L} = \bar \psi[\gamma ^\mu (i\partial _\mu - eA_\mu) - m]\psi .
\label{eq:DIRACLD}
\end{equation}
A variation with respect to $\bar \psi$ yields the corresponding
Euler-Lagrange equation
\begin{equation}
\gamma ^\mu (i\partial _\mu - eA_\mu)\psi = m\psi .
\label{eq:DIRACEQ}
\end{equation}
As stated in section 2, the dimension of a Lagrangian density is
$[L^{-4}]$. Therefore,
the dimension of $\psi $ is $[L^{-3/2}]$ and the Dirac 4-current
\begin{equation}
j^\mu = \bar \psi\gamma ^\mu \psi ,
\label{eq:DIRACJMU}
\end{equation}
satisfies the required dimension and the
continuity equation $(\!\!~\ref{eq:CONTINUITY})$ (see [7], p. 9).
Thus, the density is the 0-component of $(\!\!~\ref{eq:DIRACJMU})$
\begin{equation}
\rho _{Dirac} = \psi ^\dagger \psi .
\label{eq:DIRACRHO}
\end{equation}
Substituting the Dirac Lagrangian density $(\!\!~\ref{eq:DIRACLD})$
into the general formula $(\!\!~\ref{eq:HDENSITY})$,
one obtains the Dirac Hamiltonian density
\begin{equation}
{\mathcal H} = \psi ^\dagger [\mbox {\boldmath $\alpha $}
\cdot (-i {\mathbf \nabla } -e {\mathbf A}) +
\beta m + eV]\psi ,
\label{eq:DIRACHD}
\end{equation}
The density $\psi ^\dagger \psi $ can be
factored out from $(\!\!~\ref{eq:DIRACHD})$ and the expression
enclosed within the square brackets is the
Dirac Hamiltonian written as a differential operator.
Its substitution into $(\!\!~\ref{eq:HPSI})$ yields
the well known Dirac quantum mechanical equation
\begin{equation}
i\frac {\partial \psi}{\partial t}
 = [\mbox {\boldmath $\alpha $}
\cdot (-i {\mathbf \nabla } -e {\mathbf A}) +
\beta m + eV]\psi .
\label{eq:DIRACHOP}
\end{equation}

It is also interesting
to note that due to the linearity of the Dirac Lagrangian
density $(\!\!~\ref{eq:DIRACLD})$ with respect to $\dot \psi $,
the Dirac Hamiltonian density $(\!\!~\ref{eq:DIRACHD})$
as well as the Dirac Hamiltonian {\em do not contain a derivative
of $\psi $ with
respect to time.} Hence, $(\!\!~\ref{eq:DIRACHOP})$ is an explicit
first order differential equation.
It is easily seen that $(\!\!~\ref{eq:DIRACHOP})$ agrees completely
with the Euler-Lagrange equation $(\!\!~\ref{eq:DIRACEQ})$
of the Dirac field. It follows that Requirement 1 which is
written near the end of section 2 is satisfied.

A Hilbert space can be constructed from the eigenfunctions obtained
as solutions of the Dirac equation $(\!\!~\ref{eq:DIRACHOP})$.
Here the inner product of the Hilbert space is based on the density
of the Dirac function $(\!\!~\ref{eq:DIRACRHO})$. The eigenfunctions
of the Hamiltonian are used for building an orthonormal basis
\begin{equation}
\delta_{ij} = \int \psi ^\dagger _i \psi _j\,d^3x.
\label{eq:HIJDIRAC}
\end{equation}
Now, the form of an energy eigenfunction is
\begin{equation}
\psi ({\bf x},t) = e^{-iEt} \chi({\bf x}).
\label{eq:EEIGENFUNCTION}
\end{equation}
This form enables a construction of a Hilbert space based on
$e^{-iEt} \chi({\bf x})$ (the Schroedinger picture) {\em or}
on $\chi({\bf x})$ (the Heisenberg picture). Here, in the
Heisenberg picture, wave functions of the Hilbert space
are time independent.

As is well known, the non-relativistic limit of the Dirac equation
agrees with the Pauli equation of a spinning electron (see [7], pp. 10-13).
Hence, in accordance with the discussion presented in the first section,
the Dirac relativistic quantum mechanical equation
is consistent with the constraint imposed by the lower
rank theory of the non-relativistic quantum mechanical equations.
A related aspect of this constraint is the density
represented by the Dirac wave function $(\!\!~\ref{eq:DIRACRHO})$.
Indeed, in the non-relativistic limit of Dirac's density,
$(\!\!~\ref{eq:DIRACRHO})$
reduces to the product of the "large" components of Dirac's $\psi $
(see [7], pp. 10-13). Hence, $(\!\!~\ref{eq:DIRACRHO})$ agrees
with the density of the Pauli-Schroedinger equations $\Psi^\dagger \Psi $.
This agreement also proves the compatibility of the Hilbert space
of the Pauli-Schroedinger equations with that of the
non-relativistic limit of the Dirac equation.

Beside the satisfactory status of the Dirac theory,
this equation has an extraordinary success in describing experimental
results of electrons and muons in general and in atomic
spectroscopy in particular. Moreover, experiments of very high energy prove
that quarks are spin-1/2 particles. In particular, high energy experimental
data are consistent with the {\em point-like} nature of electrons, muons
and quarks (see [8], pp. 271, 272; [9], p. 149). Hence, the Dirac
equation satisfies item 3 of section 2.

\vglue 0.66666in
\noindent
{\bf 4. Lagrangian Density of Second Order Equations of Motion}
\vglue 0.33333in

This section discusses second order quantum equations of motion (denoted
here by SOE) which are derived from a Lagrangian density.
The presentation is analogous to that of the
previous section where the Dirac equation is discussed. The
analysis concentrates on terms containing the highest order derivatives.
Thus, the specific form of terms containing
lower order derivatives is not written explicitly
and all kinds of these terms are denoted by the acronym for Low Order
Terms $LOT$. Second order quantum differential
equations are derived from Lagrangian densities of
the following form:
\begin{equation}
{\mathcal L} = \phi ^*_{,\mu }\phi _{,\nu }g^{\mu \nu } + LOT.
\label{eq:LDEQ2}
\end{equation}
This form of the Lagrangian density is used for
the KG (see [3], p. 38) and the Higgs (see [4], p. 715)
fields.

Applying the Euler-Lagrange variational principle to
the Lagrangian density $(\!\!~\ref{eq:LDEQ2})$ one obtains a second
order differential equation that takes the following form
\begin{equation}
g^{\mu \nu }\partial _\mu \partial _\nu \phi = LOT.
\label{eq:KGEQ2}
\end{equation}

Here, unlike the case of the Dirac field, the dimension of $\phi $
is $L^{-1}$. Hence, in order to satisfy dimensional requirements, the
expression for density must contain a derivative with respect to
a coordinate. Thus, the 4-current takes the following form
(see [3], p. 40; [10], p. 199)
\begin{equation}
j_\mu = i(\phi ^* \phi _{,\mu } - \phi ^*_{,\mu } \phi) + LOT
\label{eq:KGJ}
\end{equation}
and the density is
\begin{equation}
\rho  = i(\phi ^* \dot \phi  - \dot \phi ^* \phi) + LOT.
\label{eq:KGDENSITY}
\end{equation}
The left hand side of $(\!\!~\ref{eq:KGJ})$ is a 4-vector. Therefore,
$\phi $ of SOE is a Lorentz scalar.

Using the standard method $(\!\!~\ref{eq:HDENSITY})$,
one finds that the Hamiltonian density
takes the following form (see [3], p. 38; [10], p. 198)
\begin{equation}
{\mathcal H} = \dot \phi ^* \dot \phi + (\nabla \phi ^*)\cdot (\nabla \phi)
+ LOT.
\label{eq:KGHDENSITY}
\end{equation}

An analysis of these expressions
shows that, unlike the case of the Dirac equation,
SOE theories encounter problems. Some of these problems are listed below.

\begin{itemize}
\item[{a.}] One cannot obtain a differential operator representing
the Hamiltonian. Indeed, the highest order time derivative of the
SOE density $(\!\!~\ref{eq:KGDENSITY})$
is {\em anti-symmetric}
with respect to $\dot \phi ^*,\dot \phi $ whereas the
corresponding term of the
Hamiltonian density $(\!\!~\ref{eq:KGHDENSITY})$ is {\em symmetric} with
respect to these functions (see [11], section 3, which discusses
the KG equation). Hence, in the case of SOE theories,
one cannot use method $A$ of section 2
for constructing a Hilbert space for the system.

\item[{b.}] The density associated with the wave function $\phi $
is an indispensable element of the Hilbert space.
The dependence of the SOE density $(\!\!~\ref{eq:KGDENSITY})$
on time-derivatives proves that a SOE Hilbert space {\em is built on
functions of the four space-time coordinates} $x^\mu $.
Hence, SOE cannot use
the Heisenberg picture where the functions of the Hilbert space are
time independent $\psi _H = \psi _S(t_0)$ (see [3], p. 7).

\item[{c.}] In the Schroedinger theory $\Psi ^* \Psi$ represents density.
Therefore, like the case of the Dirac field,
the dimension of this $\Psi $ is $[L^{-3/2}]$. On the
other hand, the dimension of the SOE function $\phi $ is $[L^{-1}]$.
Therefore, the nonrelativistic limit of SOE theories is inconsistent
with the Schroedinger theoretical structure.

\item[{d.}] Unlike the Dirac Hamiltonian, which is independent of
time-derivatives of $\psi $, the SOE Hamiltonian density has a term
containing the {\em bilinear product}
$\dot \phi ^* \dot \phi $. Hence, it is not clear how
a SOE analogue of the fundamental quantum mechanical equation
$(\!\!~\ref{eq:HPSI})$ can be created. Moreover, it should be
proved that this {\em first order implicit nonlinear differential
equation} is consistent with the corresponding
{\em second order explicit differential equation} $(\!\!~\ref{eq:KGEQ2})$
of SOE, as stated by requirement 1
which is formulated near the end of section 2. Without substantiating
the validity of the Hamiltonian,
SOE theories violate a constraint imposed by a lower rank theory
which is explained in the lines that precede $(\!\!~\ref{eq:EMTENSOR})$.

\item[{e.}] Some SOE theories apply to {\em real } fields (see [3],
p. 26; [4], p. 19 etc.).
New problems
arise for these kinds of physical objects. Indeed, density cannot be
defined for these particles (see [12], pp. 41-43). Moreover, a massive
particle may be at rest. In this case its amplitude should be independent
of time. But a {\em real} wave function has no phase. Therefore,
in the case of a motionless real particle, {\em the
time-derivative of its wave function
vanishes identically}. For this reason, its physical
behavior cannot be described by a differential equation with respect
to time. Thus, a real SOE particle {\em cannot} be described by the
SOE equation of motion $(\!\!~\ref{eq:KGEQ2})$ and it cannot have
a Hamiltonian.

\item[{f.}] Another problem arises for a charged SOE particle. As
stated in item $a$ above, this particle cannot have a differential
operator representing the Hamiltonian. Hence, method $A$, discussed
near $(\!\!~\ref{eq:HPSIE})$-$(\!\!~\ref{eq:HPSI})$, cannot be
used for a Hilbert space construction.
Moreover, the inner product of a time-dependent Hilbert space
is destroyed in the case of an external charge that approaches
a charged SOE particle (see [13], pp. 59-61). Hence, method $B$
does not hold either. It follows that a charged SOE particle has no
Hamiltonian. Therefore, a charged SOE particle does not satisfy a
constraint imposed by a lower rank theory.

\end{itemize}

This discussion points out theoretical difficulties of SOE fields.
The experimental side responds accordingly. Point 3 of section 2
is useful for evaluating the data. Thus, a
field $\psi (x^\mu )$ used in a Lagrangian density describes
{\em an elementary point-like particle}. It turns out that
as of today, no scalar pointlike particle has been detected.

In the history of physics, the three
$\pi $-mesons have been regarded as KG
particles and the electrically neutral $\pi ^0 $ member of this triplet
was regarded
as a Yukawa particle, namely, a real (pseudo) scalar KG particle.
However, it has already been established
that $\pi $-mesons are not elementary pointlike
particles but composite particles made of $q\bar q$ and they occupy
a nonvanishing spatial volume. Thus, as of today,
there is no experimental support for an SOE particle. The
theoretical and experimental SOE problems mentioned above are
regarded seriously here. On the basis of the foregoing
analysis, it is predicted here that no
genuine elementary SOE particle will be detected.
A special case is the
following statement: a genuine Higgs particle will not be detected.

\vglue 0.66666in
\noindent
{\bf 5. Irreducible Representations of the Poincare Group}
\vglue 0.33333in

The significance of
Wigner's analysis of the irreducible representations of the
Poincare group (see [14]; [15], pp. 44-53; [16], pp. 143-150)
is described by the following words: "It is difficult to overestimate
the importance of
this paper, which will certainly stand as one of the great
intellectual achievements of our century" (see [16], p. 149).
Wigner's work shows that there are two physically
relevant classes of irreducible representations of the Poincare group.
One class is characterized by a mass
$m>0$ and a spin $s$. The second class consists of cases where
the self mass $m=0$, the energy $E>0$ and two values of helicity.
(Helicity is the projection of the particle's spin in the direction of
its momentum.) Two values of helicity $\pm s$ correspond to a spin $s$.
Thus, each massive particle makes a
basis for a specific irreducible representation that
is characterized by the pair of values $(m,s)$. A massless particle
(like the photon) has a zero self mass, a finite energy
and two values of helicity (for a photon, the helicity is $\pm 1$).

A result of this analysis is that a system that is stable for a long
enough period of time is a basis for an irreducible
representation of the Poincare
group (see [15], pp. 48-50). Let us take a photon. Cosmic photons
are detected by measuring devices on earth after traveling
in space for a very
very long time, compared to the duration of an electromagnetic interaction.
Therefore, photons must belong to a unique irreducible representation of the
Poincare group. This conclusion is inconsistent with the idea of
Vector Meson Dominance (VMD). VMD regards the photon as a linear
combination of a massless real photon {\em and} a massive vector meson.
(For a presentation of VMD see [9], pp. 296-303; [17].)

The VMD idea has been suggested in order to explain experimental
results of scattering of energetic photons on nucleons. The main points
of the data are:
\begin{itemize}
\item[{i.}] The overall charge of a proton is $+e$ whereas the
overall charge of a neutron vanishes. Therefore, charge constituents
of a proton and a neutron are different.
\item[{ii.}] In spite of the data of the previous item, interaction of
a hard photon with a proton is nearly the same as its interaction with
a neutron.
\end{itemize}
The theoretical analysis of Wigner's work shows that VMD is
unacceptable. Other
inconsistencies of VMD with experimental data have also been published [18].
This state of affairs means that the currently accepted Standard Model
has no theoretical explanation for the photon-nucleon interaction. This
point is implicitly recognized by the PACS category of VMD which
does not belong to a theoretical PACS class. Thus, on July 2009, VMD
is included in the class of "Other models for strong interactions".
Hence, the Standard Model does not provide a theoretical
explanation for the scattering data of hard photons on nucleons.

\vglue 0.66666in
\noindent
{\bf 6. Duality Transformations of Electromagnetic Fields}
\vglue 0.33333in

Electromagnetic fields travel in vacuum at the speed of light.
Therefore, the associated particle, namely - the photon, is massless.
For this reason, it cannot be examined in a frame where it is
motionless. This result means that the argument of point $e$ of
section 4 does not hold for electromagnetic fields.
It follows that, unlike the wave function of a massive
particle, electromagnetic fields can be described by a Lagrangian
density that depends on {\em real functions}.
This well known
fact is another aspect of the inherent difference between
massive and massless particles, which has been obtained by Wigner
and discussed in the previous section.

Thus, the system consists of electromagnetic fields whose
equations of motion (Maxwell equations) are derived from
a Lagrangian density and charge carrying massive particles whose equation
of motion (the Lorentz force) is derived from a classical
Lagrangian. Below, this theory is called
ordinary electrodynamics. All quantities are described
by real functions. The action of the system is
(see [2], p. 75)
\begin{equation}
S = -\int m \sqrt {1-v^2} dt -
\int A_\mu j_{(e)}^\mu d^4 x -
\frac {1}{16\pi }\int F_{\mu \nu}F^{\mu \nu}d^4 x.
\label{eq:EMACTION}
\end{equation}
where the subscript $_{(e)}$ indicates that $j^\mu $ is a current of
electric charges. $A_\mu $ denotes the 4-potential of the electromagnetic
fields and $F^{\mu \nu }$ is the corresponding fields tensor
\begin{equation}
F_{\mu \nu } = A_{\nu ,\mu } - A_{\mu ,\nu }.
\label{eq:POT}
\end{equation}
The explicit form of this tensor is
\begin{equation}
F^{\mu \nu } = \left(
\begin{array}{cccc}
0   & -E_x & -E_y & -E_z \\
E_x &  0   & -B_z &  B_y \\
E_y &  B_z &   0  & -B_x \\
E_z & -B_y &  B_x &  0
\end{array}
\right).
\label{eq:FE}
\end{equation}
The foregoing expressions enable one to derive Maxwell equations
(see [2], pp. 78, 79 and 70, 71)
\begin{equation}
F^{\mu \nu }_{\;\,,\nu } = -4\pi j^\mu _{(e)} ;\;\;\;
F^{*\mu \nu }_{\;\;\;,\nu } = 0.
\label{eq:MAXWELL}
\end{equation}
Here $F^{*\mu \nu }$ is the dual tensor of $F^{\mu \nu }$
\begin{equation}
F^{*\mu \nu } = \left(
\begin{array}{cccc}
0   & -B_x & -B_y & -B_z \\
B_x &  0   &  E_z & -E_y \\
B_y & -E_z &   0  &  E_x \\
B_z &  E_y & -E_x &  0
\end{array}
\right).
\label{eq:FM}
\end{equation}
These tensors satisfy the following relation
\begin{equation}
F^{*\mu \nu } = \frac {1}{2}\varepsilon ^{\mu \nu \alpha \beta }
F_{\alpha \beta },
\label{eq:DUE}
\end{equation}
where $\varepsilon ^{\mu \nu \alpha \beta }$ is the completely
antisymmetric unit tensor of the fourth rank.

The Lorentz force, which describes the motion of a charged particle,
is obtained from a variation of the particle's
coordinates (see [2], pp. 49-51)
\begin{equation}
ma_{(e)}^\mu = eF^{\mu \nu }v_\nu .
\label{eq:LOR}
\end{equation}

The foregoing expressions describe the well established
theoretical structure of ordinary electrodynamics.
Let us see the results of introducing duality transformations.
Duality transformations (also called duality rotations by $\pi /2$)
of electromagnetic fields take the following
form (see [19], pp. 252, 551; [20], 1363)
\begin{equation}
\bf E \rightarrow \bf B,\;\;\;\bf B \rightarrow -\bf E.
\label{eq:DUALITY1}
\end{equation}
These transformations can be put into the following tensorial form
\begin{equation}
F^{\mu \nu } \rightarrow F^{*\mu \nu };\;\;\;
F^{*\mu \nu } \rightarrow -F^{\mu \nu }.
\label{eq:DUALITY2}
\end{equation}

An examination of the {\em homogeneous} Maxwell equations
\begin{equation}
F^{\mu \nu }_{\;\,,\nu }=0;\;\;\;
F^{*\mu \nu }_{\;\;\;,\nu }=0,
\label{eq:HOMOGENOUS}
\end{equation}
proves that they are invariant under the duality transformations
$(\!\!~\ref{eq:DUALITY2})$. On the other hand, an inequality
is obtained for the inhomogeneous Maxwell equation
\begin{equation}
F^{*\mu \nu }_{\;\;\;,\nu } \neq -4\pi j^\mu _{(e)}.
\label{eq:MAXWELLDU1}
\end{equation}
This problem can be settled by the introduction of the notion of
magnetic monopoles (called briefly monopoles). Thus, duality
transformations of the electromagnetic fields
$(\!\!~\ref{eq:DUALITY2})$ are augmented by the following
transformation that relates charges and monopoles
\begin{equation}
e \rightarrow g;\;\;\;g\rightarrow -e,
\label{eq:EG}
\end{equation}
where $g$ denotes the monopole strength.

Two things are established at this point:
\begin{itemize}
\item[{1.}] The theoretical foundation of ordinary electrodynamics
$(\!\!~\ref{eq:EMACTION})$, and its equations of motion
$(\!\!~\ref{eq:MAXWELL})$ and $(\!\!~\ref{eq:LOR})$.
\item[{2.}] The mathematical form of duality transformations
$(\!\!~\ref{eq:DUALITY2})$ and $(\!\!~\ref{eq:EG})$.
\end{itemize}
Now, a theory for a system of monopoles and electromagnetic fields
(called below monopole electrodynamics) is
obtained from the application of duality transformations to
ordinary electrodynamics. The action principle of this system is
\begin{equation}
S = -\int m \sqrt {1-v^2} dt -
\int A_{(m)\mu} j_{(m)}^\mu d^4 x -
\frac {1}{16\pi }\int F^*_{(m)\mu \nu}F_{(m)}^{*\mu \nu}d^4 x,
\label{eq:EMACTIONM}
\end{equation}
where the subscript $(m)$ denotes that the quantities pertain to
monopole electrodynamics. Here the fields are derived from a
4-potential
\begin{equation}
F^*_{(m)\mu \nu } = A_{(m)\nu ,\mu } - A_{(m)\mu ,\nu },
\label{eq:POTM}
\end{equation}
which is analogous to $(\!\!~\ref{eq:POT})$. Maxwell equations of
monopole electrodynamics are
\begin{equation}
F_{(m)\;\,,\nu }^{*\;\;\;\;\mu \nu } = -4\pi j^\mu _{(m)} ;\;\;\;
F^{\;\;\;\;\;\mu \nu }_{(m)\,,\nu } = 0
\label{eq:MAXWELLM}
\end{equation}
and the Lorentz force is
\begin{equation}
ma_{(m)}^\mu = gF_{(m)}^{*\mu \nu }v_\nu .
\label{eq:LORM}
\end{equation}

Thus, we have two theories for two distinct systems:
ordinary electrodynamics for a system of charges and fields
and monopole electrodynamics for a system of monopoles and fields.
The first system does not contain monopoles and the second system does
not contain charges.
The problem is to find the form of a unified theory that
describes the motion of charges, monopoles and fields. Below,
such a theory is called a charge-monopoly theory. The charge-monopole
theory is a higher rank theory whose domain of validity includes those
of ordinary electrodynamics and of monopole electrodynamics
as well. On
undertaking this assignment, one may examine two postulates:
\begin{itemize}
\item[{1.}] Electromagnetic fields of ordinary electrodynamics
are identical to electromagnetic fields of monopole electrodynamics.
\item[{2.}] The limit of the charge-monopole theory for a system
that does not contain monopoles agrees with ordinary electrodynamics
and limit of the charge-monopole theory for a system
that does not contain charges agrees with monopole electrodynamics.
\end{itemize}
It turns out that these postulates are mutually contradictory.

A charge-monopole theory that relies (implicitly) on the first
postulate has been published by Dirac many years ago [21,22].
(Ramifications of Dirac monopole theory can be found in the
literature [20].)
This theory shows the need to define physically unfavorable
irregularities along strings. Moreover,
the form of its limit that applies
to a system of monopoles without charges is inconsistent with the
theory of
monopole electrodynamics, which is derived above from the duality
transformations. Therefore, it does not satisfy the constraint
imposed by a lower rank theory. The present experimental situation is that
in spite of a long search,
there is still no confirmation of the existence of a Dirac monopole
(see [23], pp. 1209).

The second postulate was used for constructing a different charge-monopole
electrodynamics [24,25]. This postulate guarantees that the
constraints imposed by the two lower rank theories are satisfied.
Moreover,
this theory does not introduce new irregularities into electrodynamics.
Thus, it is called below regular charge-monopole theory.
The following statements describe important results of the
regular charge-monopole theory:
The theory can be derived from an action principle, whose
limits take the form of
$(\!\!~\ref{eq:EMACTION})$ and $(\!\!~\ref{eq:EMACTIONM})$, respectively.
Charges do not interact with bound fields of monopoles; monopoles do not
interact with bound fields of charges; radiation fields
(namely, photons) of the systems
are identical and charges as well as monopoles interact with them.
Another result of this theory is that the size of an elementary monopole $g$
is a free parameter. Hence, the theory is relieved from the huge
and unphysical Dirac's monopole size $g^2 = 34.25$.

The regular charge-monopole theory is constructed on the basis
of the second postulate. This point means that it is not guided by
new experimental data. However,
it turns out that it explains the important
property of hard photon-nucleon interaction which is mentioned in
the previous section. Indeed, just assume that quarks carry
a monopole and
postulate that the elementary monopole unit $g$ is much larger then the
electric charge $e$ (probably $|g|\simeq 1$). This property means that
photon-quark interaction depends mainly on monopoles and that
the photon interaction with the quarks' electric charge is a
small perturbation. Therefore,
the very similar results of photon-proton and photon-neutron scattering
are explained. (Note also that all baryons have a core which
carries three units of magnetic charge that attracts the three
valence quarks. The overall magnetic charge of a hadron vanishes.)
Other kinds of experimental support for the
regular charge-monopole theory have been published elsewhere [26].

\vglue 0.66666in
\noindent
{\bf 7. Concluding Remarks}
\vglue 0.33333in

This work relies on the main assumption of theoretical physics which
states that results derived from physically relevant mathematical
structures are expected to fit experimental data [27]. Three
well known mathematical structures are used here: the variational
principle, Wigner's analysis of the irreducible representations
of the Poincare group and duality transformations of electromagnetic
fields.

The paper explains and uses three points which are either new or
at least lack an adequate discussion in textbooks.
\begin{itemize}
\item[{1.}] Constraints are imposed by a lower rank theory on properties of
the corresponding limit of a higher rank theory (see a discussion in
the Introduction).
\item[{2.}] The need to prove consistency between the Euler-Lagrange equation
obtained from a Lagrangian density and the quantum mechanical
equation $i\partial \psi /\partial t = H\psi $ which holds for the
corresponding Hamiltonian.
\item[{3.}] The field function $\psi (x^\mu )$ describes an elementary
pointlike particle (see the discussion near $(\!\!~\ref{eq:PSI})$).
\end{itemize}
Points 1 and 2 are useful for a theoretical evaluation of the
acceptability of specific physical ideas. Point 3 is useful for finding an
experimental support for these ideas.

The main results of the analysis presented in this work are as follows:
Dirac equation is theoretically consistent and has an enormous
experimental support. Second order quantum mechanical equations
(like the Klein-Gordon and the Higgs equations) suffer from many
theoretical problems and have no experimental support. ($\pi $-mesons
are not pointlike, therefore, they are not genuine Klein-Gordon
particles.) Real fields cannot be used for a description
of massive particles.
The idea of Vector Meson Dominance is inconsistent with
Wigner's analysis of the irreducible representations of the Poincare
group. Therefore, VMD is unacceptable and the Standard Model has no
theoretical explanation for the data of a scattering process of
an energetic photon on nucleon. Monopole theories that
introduce irregularities along strings are inconsistent with point
1 of this section and have no experimental support. The regular
charge monopole theory [24-26] is consistent with point 1 and has
experimental support.


\newpage
References:
\begin{itemize}
\item[{*}] Email: elicomay@post.tau.ac.il  \\
\hspace{0.5cm}
           Internet site: http://www.tau.ac.il/$\sim $elicomay

\item[{[1]}] F. Rohrlich, {\em Classical Charged Particles},
(Addison-wesley, Reading Mass, 1965).
\item[{[2]}] L. D. Landau and E. M. Lifshitz, {\em The Classical
Theory of Fields} (Elsevier, Amsterdam, 2005).
\item[{[3]}] J. D. Bjorken and S.D. Drell, {\em Relativistic Quantum
Fields} (McGraw-Hill, New York, 1965).
\item[{[4]}] M. E. Peskin and D. V. Schroeder, {\em An Introduction to
Quantum Field Theory} (Addison-Wesley, Reading, Mass., 1995).
\item[{[5]}] L. D. Landau and E. M. Lifshitz, {\em Quantum Mechanics}
(Pergamon, London, 1959).
\item[{[6]}] L. I. Schiff, {\em Quantum Mechanics} (McGraw-Hill, New York,
1955).
\item[{[7]}] J. D. Bjorken and S.D. Drell, {\em Relativistic Quantum
Mechanics} (McGraw-Hill, New York, 1964).
\item[{[8]}] D. H. Perkins, {\em Introduction to High Energy Physics}
(Addison-Wesley, Menlo Park, CA, 1987).
\item[{[9]}] H. Frauenfelder and E. M. Henley, {\em Subatomic Physics},
(Prentice Hall, Englewood Cliffs, 1991).
\item[{[10]}] W. Pauli and V. Weisskopf, Helv. Phys. Acta, {\bf 7}, 709 (1934).
English translation: A. I. Miller {\em Early Quantum
Electrodynamics} (University Press, Cambridge, 1994). pp. 188-205.
(In the text, page references apply to the English translation.)
\item[{[11]}] E. Comay, Apeiron {\bf 12}, 26 (2005).
\item[{[12]}] V. B. Berestetskii, E. M. Lifshitz and L. P. Pitaevskii,
{\em Quantum Electrodynamics} (Pergamon, Oxford, 1982).
\item[{[13]}] E. Comay, Apeiron {\bf 14}, 50 (2007).
\item[{[14]}] E. Wigner, Ann. Math. {\bf 40} 149 (1939).
\item[{[15]}] S. S. Schweber, {\em An Introduction to Relativistic
Quantum Field Theory}, (Harper \& Row, New York, 1964).
\item[{[16]}] S. Sternberg, {\em Group Theory and Physics}
(Cambridge University Press, Cambridge, 1994).
\item[{[17]}] T. H. Bauer, R. D. Spital, D. R. Yennie and F. M. Pipkin,
Rev. Mod. Phys. {\bf 50}, 261 (1978).
\item[{[18]}] E. Comay, Apeiron {\bf 10}, 87 (2003).
\item[{[19]}] J. D. Jackson, Classical Electrodynamics (John Wiley,
New York, 1975).
\item[{[20]}] P. Goddard and D. I. Olive, Rep. Prog. Phys.,
{\bf 41}, 1357 (1978).
\item[{[21]}] P. A. M. Dirac, Proc. Royal Soc. {\bf A133}, 60 (1931).
\item[{[22]}] P. A. M. Dirac, Phys. Rev., {\bf {74}}, 817 (1948).
\item[{[23]}] C. Amsler et al., Phys. Lett. {\bf B667}, 1 (2008).
\item[{[24]}] E. Comay, Nuovo Cimento, {\bf 80B}, 159 (1984).
\item[{[25]}] E. Comay, Nuovo Cimento, {\bf 110B}, 1347 (1995).
\item[{[26]}]
E. Comay, published in Has the Last Word Been Said on Classical
Electrodynamics? Editors: A. Chubykalo, V. Onoochin,
A. Espinoza, and R. Smirnov-Rueda (Rinton Press, Paramus, NJ,
2004). (The article's title is {\em A Regular Theory of Magnetic Monopoles
and Its Implications.})
\item[{[27]}] E. P. Wigner, Comm. in Pure and Appl. Math. {\bf 13}, 1 (1960).

\end{itemize}


\end{document}